\documentclass[journal=jacsat,manuscript=article]{achemso}

\usepackage{chemformula} 
\usepackage[T1]{fontenc} 
\usepackage{graphicx}
\usepackage{soul} 

\usepackage{xcolor}
\usepackage{siunitx}
\usepackage{xr} 

\DeclareRobustCommand*{\inlinecite}[1]{%
	\begingroup
	\romannumeral-`\x 
	\setcitestyle{numbers}%
	\cite{#1}%
	\endgroup
}

\author{Xing He, Kairi Masuda, Aaron M. Schankler, Songsong Zhou, Andrew M. Rappe}
\affiliation[University of Pennsylvania]
{Department of Chemistry, University of Pennsylvania, Philadelphia, Pennsylvania 19104-6323, USA }
\email{rappe@sas.upenn.edu}

\title[An \textsf{achemso} demo]
 {Uncovering the Structural, Electronic, and Electrostatic Landscapes of SrTiO$_3$ Grain Boundaries via Grand Canonical Monte Carlo with a Machine-Learning Interatomic Potential}

\abbreviations{IR,NMR,UV}

\begin{document}
\hfil \date{\today} \hfil







\begin{abstract}
Grain boundaries in perovskites exhibit a rich variety of electronic, polar and magnetic properties, yet identifying their atomic structures and compositions remains challenging. Here, we combine the Grand Canonical Monte Carlo (GCMC) method with a universal machine-learning interatomic potential to explore the coupled compositional and structural phase space of $\Sigma$5(310)[001] SrTiO$_3$ grain boundaries. The resulting chemical potential phase diagram identifies ten stable structures belonging to four motif families: large Sr kite, small Sr kite, incomplete Sr large kite, and Ti-rich structures. Extending the GCMC method to the $\Sigma$5(210)[001] and $\Sigma$17(410)[001] grain boundaries reveals how grain boundary plane and inclination angle influence the competition between small- and large-kite motifs. Density functional theory calculations of the predicted $\Sigma$5(310)[001] grain boundaries show that the composition and atomic structure strongly govern the spatial distribution of conduction and valence band edge states, interfacial band alignment, and carrier localization. Four distinct electronic regimes emerge: localized electron-doped, delocalized electron-doped, insulating, and hole-attracting states. Our electrostatics analysis further reveals the coexistence of the short-range electric fields at the reconstructed GB core associated with local atomic density and bonding, and the longer-range fields outside the core arising from the grain boundary charge and carrier redistribution. These fields can also couple to local polar distortions and generate composition-dependent polarization patterns. These results establish an efficient framework for exploring oxide perovskite grain boundary structures and reveal how stoichiometry and atomic reconstruction couple to electronic structure, electrostatics, and local polarization, providing a route toward predictive grain boundary engineering in functional perovskites.

\end{abstract}

\clearpage
\section{Introduction}

Perovskite materials exhibit diverse functional properties, including piezoelectricity, magnetism, ferroelectricity, photovoltaic effects, and superconductivity \mbox{\cite{Cohen1992,PhysRevLett.109.116601,RevModPhys.77.1083,gastiasoro2020superconductivity}}. Owing to these functionalities, they have been widely used in devices such as  actuators, sensors, energy-conversion devices, and nonvolatile memory elements \mbox{\cite{doi:10.1126/science.1129564}}. In practical applications, many functional perovskite oxides are used as polycrystalline ceramics, in which grain boundaries (GBs)—interfaces between neighboring crystallites—strongly influence material properties and device performance. Although their structural cores are confined to narrow interfacial regions \mbox{\cite{Hirth1972,GB_hardening,GB_nanograin}}, GBs have significant effects on device performance. The structural, chemical, and electrostatic disturbance of GBs can give rise to emergent phenomena such as flexoelectricity, multiferroicity, and local electric fields \mbox{\cite{PhysRevLett.120.267601,shimada2015multiferroic,Meiners2020,wu2022engineering,CANTWELL20141}}. 
Recent advances have highlighted GBs not only as defects but as opportunities for enhancing material performance, for example, through textured ceramics, charged GBs, and GB engineering \mbox{\cite{doi:10.1126/science.adf6161,doi:10.1126/science.aaa4166,doi:10.1126/science.aao3691,defferriere2025grain,wang2026grain,shi2026engineering}}. Understanding the relationships among GB structure, composition, and functional behavior has therefore
become an important objective in perovskite oxide research \mbox{\cite{STO_GB,BTO_GB,gao2018atomic}}. 

However, predicting GB structures in perovskite oxides remains challenging because they are multi-elemental systems, whose composition is determined by a delicate balance between the chemical bonding energies, chemical potentials of the elements, and electrostatic interactions. Combining grand canonical Monte Carlo (GCMC) and density functional theory (DFT), that is, {\em{ab initio}} GCMC ({\em{ai}}-GCMC), provides a framework for exploring these coupled degrees of freedom. Initially applied to predict surface growth\mbox{\cite{abinitioGCMC,PhysRevB.Zhou,PhysRevLett.Zhou,qin2024unveiling}}, this approach generates stable and metastable structures based on thermodynamics with DFT accuracy, whereas other classical interatomic potentials have been used in previous GCMC simulations and other generative-type predictions such as evolutionary algorithms \mbox{\cite{ReaxFF.GCMC,GCMC_swap,Zhu2018,Chua2010,YANG2020109812}}.
Recently, this concept of $ai$-GCMC has been applied to the prediction of GB structure and composition of Al$_2$O$_3$ \cite{masuda2025ab}, treating the GB as a surface growth problem between two slabs. This strategy offers a promising route to analyzing perovskite GBs, but their extensive configurational spaces make direct DFT-based sampling computationally costly. Universal machine-learning interatomic potentials (uMLIPs), trained on datasets based on millions of DFT calculations, have proven to exhibit tens of meV/atom accuracy compared with DFT calculations, providing a new pathway to perform high-throughput searches for inorganic materials \cite{riebesell2025framework,jain2013commentary,horton2025accelerated,barroso_omat24,rhodes2025orb,fu2025learning,kim2024data}. Integrating these potentials with GCMC could therefore enable extensive searches for compositionally complex GB structures. 

In this work, we combine GCMC sampling with a uMLIP to explore the compositional and structural degrees of freedom of perovskite oxide GBs. We first applied this approach to $\Sigma$5(310)[001] GBs of SrTiO$_{3}$, a prototypical paraelectric perovskite, and the resulting stable structures, belonging to four groups of motifs (large Sr kite, small Sr kite, incomplete large Sr kite, Ti-rich), reproduce experimentally observed GB structures, supporting the applicability of this approach to complex oxide interfaces. We further investigate $\Sigma$5(210)[001] and $\Sigma$17(410)[001] GBs to examine how the GB plane and misorientation angles influence the small- and large-kite competition. DFT calculations of $\Sigma$5(310)[001] GBs reveal composition and structural-dependent electronic behavior, exhibiting four distinct electronic regimes including localized electron-doped, delocalized electron-doped, insulating, and hole-attracting states. Electrostatics analysis further reveals the coexistence of short-range GB core electric field, superimposed on the longer-range electric field arising from the space charge. These results establish direct links among GB stoichiometry, atomic structure, electronic reconstruction, electrostatics, and local polarization, providing a foundation for predictive grain boundary engineering in functional perovskite oxides.

\section{Methods}
\subsection{Theory}
In the grand canonical Monte Carlo theory, we fix the chemical potential $\mu$, volume $V$, and temperature $T$. Then, to explore the structural space, we generate GBs with the following six actions with corresponding acceptance probabilities (derivations are found in Ref.~{\inlinecite{Gibbs_2010,GCMC_frankel}}). When we attempt the move (displacement of atoms) action, the probability of accepting the move action is described as follows:
\begin{equation}
\label{eq:pmove}
P_{\rm{move}}=\mathrm{min}\{1, \: \mathrm{e}^{-\Delta U/k_{B}T}\},
\end{equation}
where $k_{B}$ is the Boltzmann constant, and $\Delta U$ is the change in total energy. 
On the other hand, the number of particles with atomic species $i$ is changed when we attempt the add and remove actions. Their probabilities of acceptance are defined as follows:
\begin{equation}
\label{eq:padd}
P_{\rm{add}}=\mathrm{min}\Big\{1, \frac{V}{(N_{i}+1)\Lambda_{i}^{3}}\: \mathrm{e}^{-(\Delta U-\mu_{i})/k_{B}T}\Big\},
\end{equation}
\begin{equation}
\label{eq:premove}
P_{\rm{remove}}=\mathrm{min}\Big\{1, \frac{N_{i} \Lambda_{i}^{3}}{V}\: \mathrm{e}^{-(\Delta U+\mu_{i})/k_{B}T}\Big\},
\end{equation}
where $\Lambda$ is the thermal de Broglie wavelength, given by $\Lambda_{i}=h/\sqrt{2\pi m_{i} k_{B}T}$, $h$ is the Planck constant, and $m_{i}$ is the mass of atomic species $i$. $N_{i}$ and $\mu_{i}$ are the total number and chemical potential of species $i$. 
For the swap (the exchange of two atom positions) action, the probability of accepting this action is described as follows:
\begin{equation}
\label{eq:swap}
P_{\rm{swap}}=\mathrm{min}\{1, \: \mathrm{e}^{-\Delta U/k_{B}T}\}.
\end{equation}
The above four actions are typically employed in GCMC and ai-GCMC simulations for surface growth predictions\mbox{\cite{abinitioGCMC,PhysRevB.Zhou,PhysRevLett.Zhou,ReaxFF.GCMC,qin2024unveiling}}. 
Furthermore, we consider a direct replacement of atoms (an exchange of atomic species by removal and addition at the same time) as a new action with the following acceptance rule\mbox{\cite{GCMC_swap}}:
\begin{equation}
\label{eq:preplace}
P_{\rm{replacement}}=\mathrm{min}\Big\{1, \: \frac{N_{i}\Lambda^{3}_{i}}{(N_{j}+1)\Lambda^{3}_{j}}\mathrm{e}^{-(\Delta U+\mu_{i}-\mu_{j})/k_{B}T}\Big\}.
\end{equation}
This action is introduced to avoid the problem that removal and addition actions are much less likely to happen after the GB region is filled. 
Moreover, the width of a GB is generally unknown. To overcome this problem, we include a grain move as a new action. In this action, entire slabs are displaced from each other to search for the optimal GB width with the following acceptance rule:
\begin{equation}
\label{eq:pgrain}
P_{\rm{grain}}=\mathrm{min}\{1, \: \mathrm{e}^{-\Delta U/k_{B}T}\}.
\end{equation}
Besides searching for the optimal width of GB, this step is also equivalent to sampling the $\gamma$ surface of the GB \cite{guo2017structural,lee2011characterization}.
In all, we consider six actions: move, add, remove, swap, replacement, and grain move actions in GCMC GB predictions.

In the $ai$-GCMC concept, density functional theory (DFT) is employed to calculate the total energy $U$ of GBs at each step after DFT structural relaxation because otherwise, the acceptance ratio can be too low. However, DFT calculation is too expensive for perovskite GB GCMC simulations because a great number of relaxation and total energy evaluations are needed. 

The uMLIP methodology has recently been further extended and can be applied to perovskite materials for the following reasons: (1) Due to the sophistication of these architectures, e.g., learning architecture with over 50 M learning parameters, the performance of this class of potentials has been significantly improved\mbox{\cite{neumann2024orbfastscalableneural,yang2024mattersim,equiformer_v2}}. (2) The variety of training datasets has also been increased recently, including sources such as open catalysts(OC)20, OC22, Alexandria, and OMat24, which includes the dataset of oxides\mbox{\cite{doi:10.1021/acscatal.0c04525,doi:10.1021/acscatal.2c05426,SCHMIDT2024101560,barroso_omat24,MPtraj_Deng2023}}. (3) Long-range interactions are naturally incorporated in the latest models employing the Transformer-attention mechanism and message passing \mbox{\cite{neumann2024orbfastscalableneural,yang2024mattersim,equiformer_v2}}. 
Therefore, we chose to employ uMLIPs for structural and compositional evolution, as well as energy evaluation of GBs, to reduce calculation cost. To ensure accuracy, we validate sampled structures from GCMC results with DFT.

\subsection{Simulation detail}
We apply the GCMC approach to $\Sigma$5(310)[001] SrTiO$_3$ GBs, with a misorientation angle of $36.9^\circ$ about the [001] axis and a (310) boundary plane.
$\Sigma$5 denotes that the volume of the coincident site lattice (CSL) between two grains is 5 times of the unit cell volume.
As illustrated in Figure S1(a), the initial structure for GCMC calculations consists of two slabs constructed from cubic SrTiO$_3$ with a lattice parameter of $a = 3.9127$\,\AA\, rotated into the desired orientations, with an inter-GB vacuum region of 5.0 \mbox{Å}.
The interfacial plane area is $\sqrt{3^2+1} \times 3.9127 \times 3.9127$\,\AA$^2$ = 48.41 \AA$^2$\ giving an initial GB region of 242.06\,\AA$^3$. In this region, we allow the move, add, remove, swap, and replacement actions. Furthermore, we set a buffer region of 10.0 \mbox{Å} on each side of the central GB region, in which we only allow atomic move actions. Atoms in the remaining outer region of each grain are fixed relative to their respective grains, and only apply grain move actions. An additional 10.0\,\AA\ vacuum spacing on each side separates the outer slab surfaces from their periodic images. The initial supercell dimension is 12.37\,\AA $\times$ 3.91 \mbox{Å} $\times$ 69.54 \mbox{Å} with a composition of Sr$_{36}$, Ti$_{36}$, and O$_{108}$. 
Throughout the simulation, we set the temperature to 1800 K, considering that $T \approx 1500 ^{\circ}$C is employed for connecting two grains by the thermal diffusion technique in the experiments\mbox{\cite{10.1063/1.3204476,Yang02012014}}. We choose five sets of chemical potentials from the stable region of bulk SrTiO$_3$, as shown in Fig.~S2. Additional simulation details are provided in the Supplemental Materials, Section~S1 (Atomic structures were visualized using VESTA \cite{momma2011vesta}). To have a closer comparison with scanning transmission electron microscopy (STEM) images, the displayed atomic sizes were chosen to scale with atomic radii, using Sr:Ti:O = 2.15 \AA : 1.47 \AA : 0.74 \AA, instead of the ionic radii.

\begin{figure}
	\includegraphics[width=0.9\textwidth]{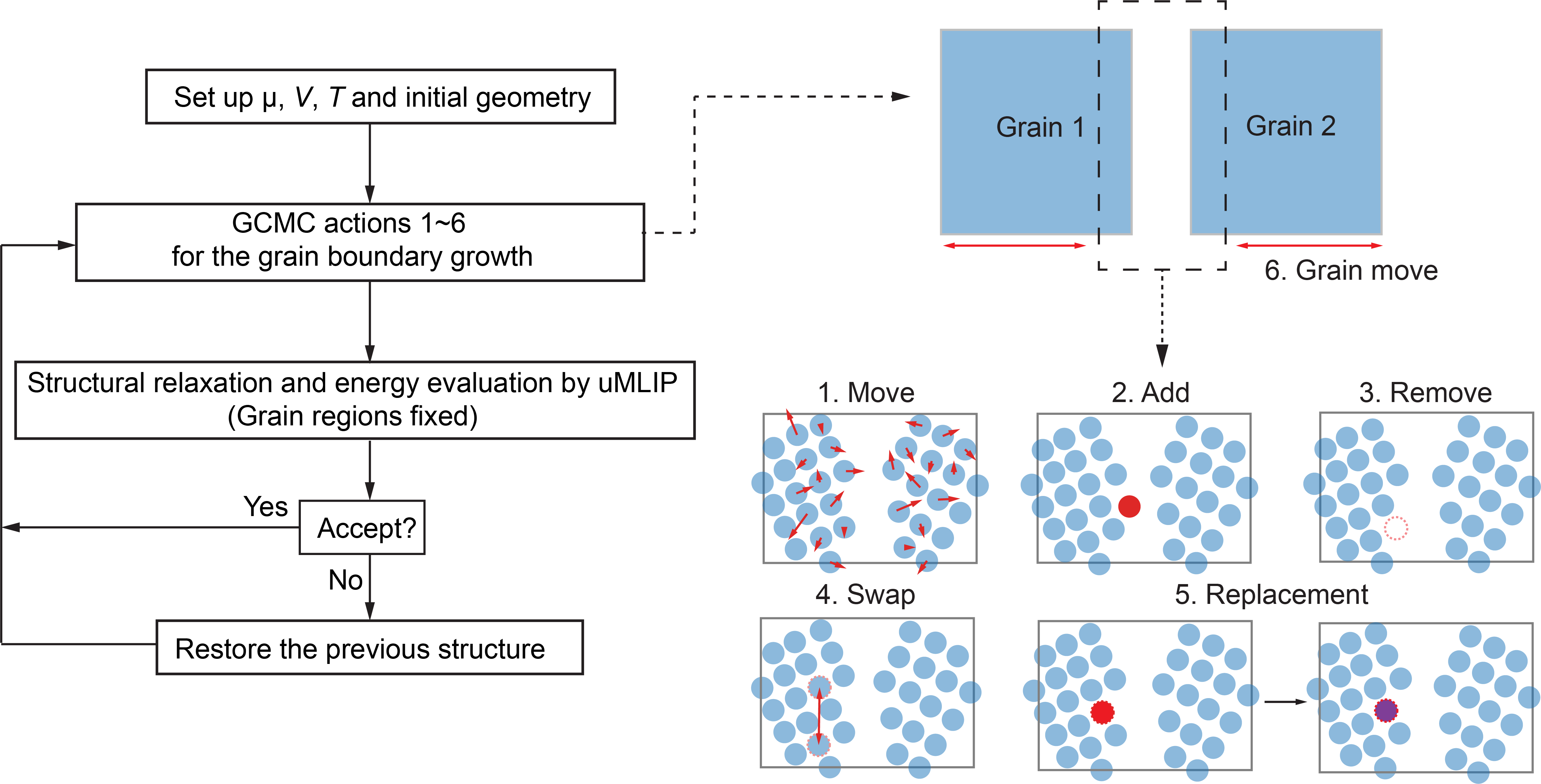}
	\caption{A schematic illustration of the GCMC framework for GB structure prediction.}
	\label{fig:GBMC_schematic}
\end{figure}

Once $\mu$, $V$, and $T$ are determined, the GCMC simulations are conducted as follows. (1) For the move action, random displacements of $-$0.1 \mbox{Å} --- $+$0.1 \mbox{Å} are applied to each atom in the GB and buffer regions. (2) For the add action, one element is chosen with an equal probability and an atom is added to the system. Instead of randomly selecting the position of the new atom and testing it directly, we include a configurational bias, which prevents the new atom from being too close ($r_{\rm{min}}$ = 1.5 \mbox{Å}) or too far ($r_{\rm{max}}$ = 2.5 \mbox{Å}) from the closest existing atom. (3) For the remove action, an atom is randomly chosen and removed from the system. (4) For the swap action, we choose two atoms randomly and exchange their positions. (5) For the replacement action, we choose one atom and replace its species with another species. (6) For the grain move action, we apply random displacements of -0.1 \mbox{Å} -- 0.1 \mbox{Å} to both grains in the $x$, $y$, and $z$ directions. We treat this action and the move action of (1) as a single action, since the acceptance rule is the same. A schematic plot of the GCMC process is summarized in Figure ~\ref{fig:GBMC_schematic}.

For the uMLIP energy evaluation and structural relaxation, we employed the Equiformer V2 86M model\mbox{\cite{equiformer_v2}} (eqV2-M), whose architecture is based on the Transformer model. The eqV2-M model includes 86.6 M parameters and is trained on the OMat24 dataset first, and fine-tuned on Materials Project and Alexandria datasets\mbox{\cite{SCHMIDT2024101560,barroso_omat24,MPtraj_Deng2023}} with the FAIR Chemistry Python module\mbox{\cite{doi:10.1021/acscatal.0c04525}}. For structural relaxation, we used force convergence thresholds of 0.05 eV/\mbox{Å} in the limited-memory BFGS (LBFGS), as implemented in the atomic simulation environment (ASE)\mbox{\cite{ase-paper}} Python module.

At each step, we accept or reject each attempted action based on the probabilities determined by the free energy change, as mentioned above. We consider this acceptance or rejection of one GCMC action as one step and conduct 1000 steps until the structure converges. Then, we conduct GCMC simulations 14 times at each set of chemical potential values to obtain sufficient samples, resulting in 14000 steps. Since we study five sets of chemical potential values, 70000 GCMC steps are generated in total. 
Following the same procedure, we also perform GCMC simulations for  $\Sigma$5(210)[001] and $\Sigma$17(410)[001] SrTiO$_3$ GBs. The $\Sigma$5(210)[001] GB shares the same CSL index and misorientation angle $36.8^{\circ}$ as $\Sigma$5(310)[001] GB, but it has a different GB plane $(210)$, as illustrated in Supplemental Materials Figure~S10. Thus, $\Sigma$5(210)[001] GB plane exhibits a larger inclination angle with the grain as $2\tan^{-1}(\frac{1}{2})\approx53.1^{\circ}$. More generally, a GB can be described by five macroscopic parameters: two parameters specifying the misorientation axis, one parameter specifying the misorientation angle, and two parameters specifying the orientation of the GB plane. Within this description, the $\Sigma$5(310)[001] and $\Sigma$5(210)[001] GBs share the same first three parameters but the orientation of the GB plane is different.
On the other hand, $\Sigma$17(410)[001] has a smaller misorientation angle as, $2\tan^{-1}(\frac{1}{4})\approx28.1^{\circ}$, and also a different GB plane $(410)$, as illustrated in Figure~S11.

\subsection{Results and discussion}
\subsubsection{Structure evolution}
\begin{figure}
	\includegraphics[width=0.6\textwidth]{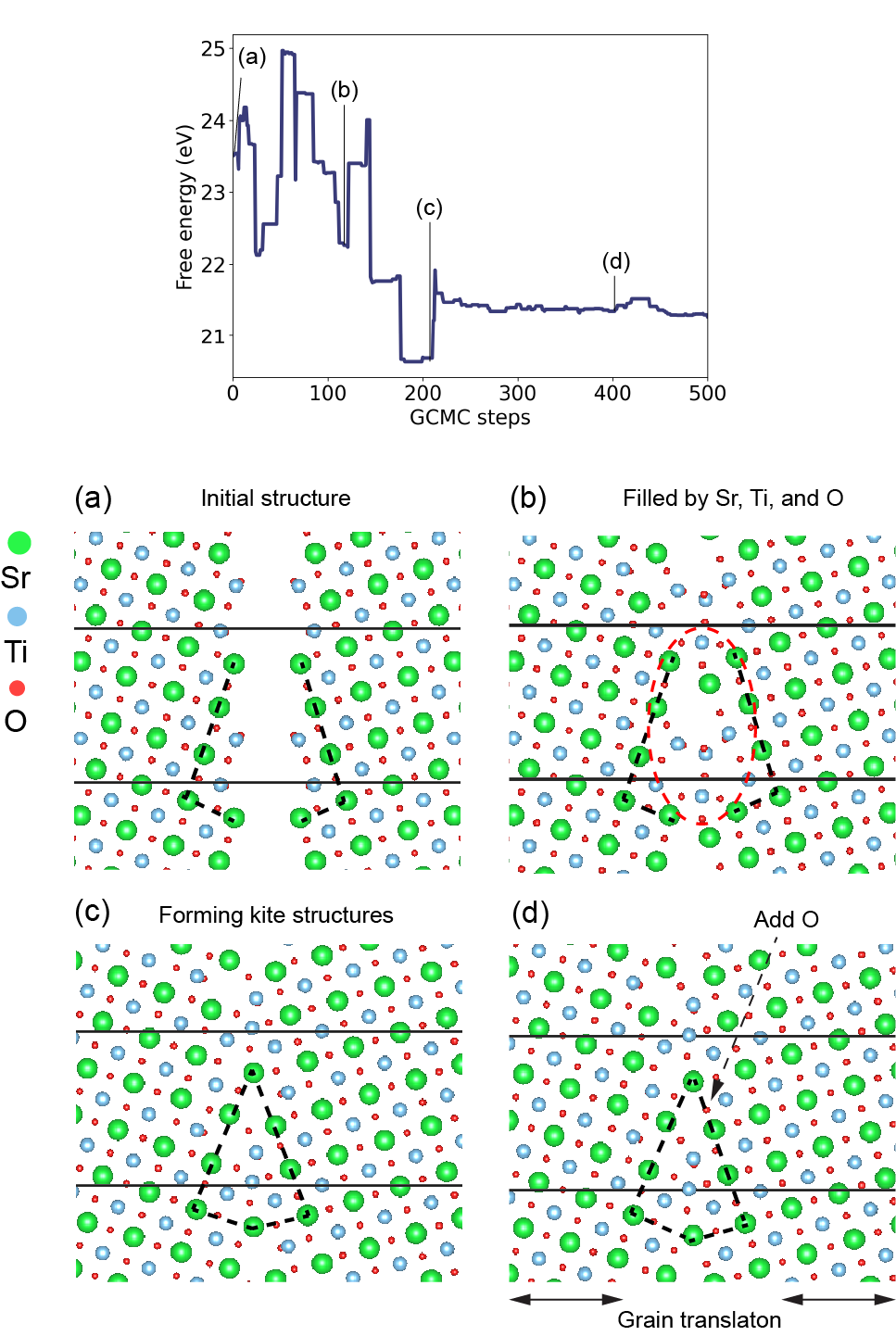}
	\caption{An example of free energy, compositional, and structural evolution of $\Sigma$5(310)[001] SrTiO$_{3}$ GBs.  (top) Free energy evolution during GCMC simulation. (a) Initial GB structure. (b) GB filled by adding Sr, Ti, and O. (c) GB kite structure formed. (d) GB structure adjustment to change composition and structure.}
	\label{fig:GB_evolution}
\end{figure}

First, we illustrate the structural evolution of a GB during a GCMC simulation to provide a clearer picture of our theory. Figure~\ref{fig:GB_evolution} and Supplemental Movie 1 show an example for a $\Sigma$5(310)[001] GB. During the simulation, we monitor the free energy $F$ at a set of certain chosen chemical potentials $\mu_i$, $F = U - \sum_i\mu_i N_i$, where $U$ is the total energy of the supercell, and $N_i$ is the number of atoms of species $i$. As shown in the upper panel of Figure~\ref{fig:GB_evolution}, $F$ initially increases as the GB region is progressively filled by adding Sr, Ti, and O atoms and then fluctuates as the system explores structural and compositional phase space (Figs.~\ref{fig:GB_evolution}(a) and (b)). The initially disordered GB subsequently develops an ordered structure containing characteristic kite motifs, accompanied by a further decrease in $F$ (Fig.~\ref{fig:GB_evolution}(c)). The GB then continues to adjust both its composition and atomic structure (Fig.~\ref{fig:GB_evolution}(d)), eventually converging toward well-defined kite-like configurations. These generated kite structures have been widely observed in experiments \mbox{\cite{yang2021determination,liao2021nanoscale,PhysRevB.78.245320}}, as discussed further below.  By repeating the GCMC sampling over a range of chemical potentials, our methodology thoroughly explores the compositional and structural phase space of SrTiO$_3$ GBs, including configurations consistent with experimental observations \cite{yang2021determination,wu2022engineering}. Because each Monte Carlo move is accepted probabilistically, the grand potential can either increase or decrease during the simulation, allowing the system to transiently sample, less energetically favorable configurations (e.g. Fig.~\ref{fig:GB_evolution}(d)). 

\begin{figure}
	\includegraphics[width=0.8\textwidth]{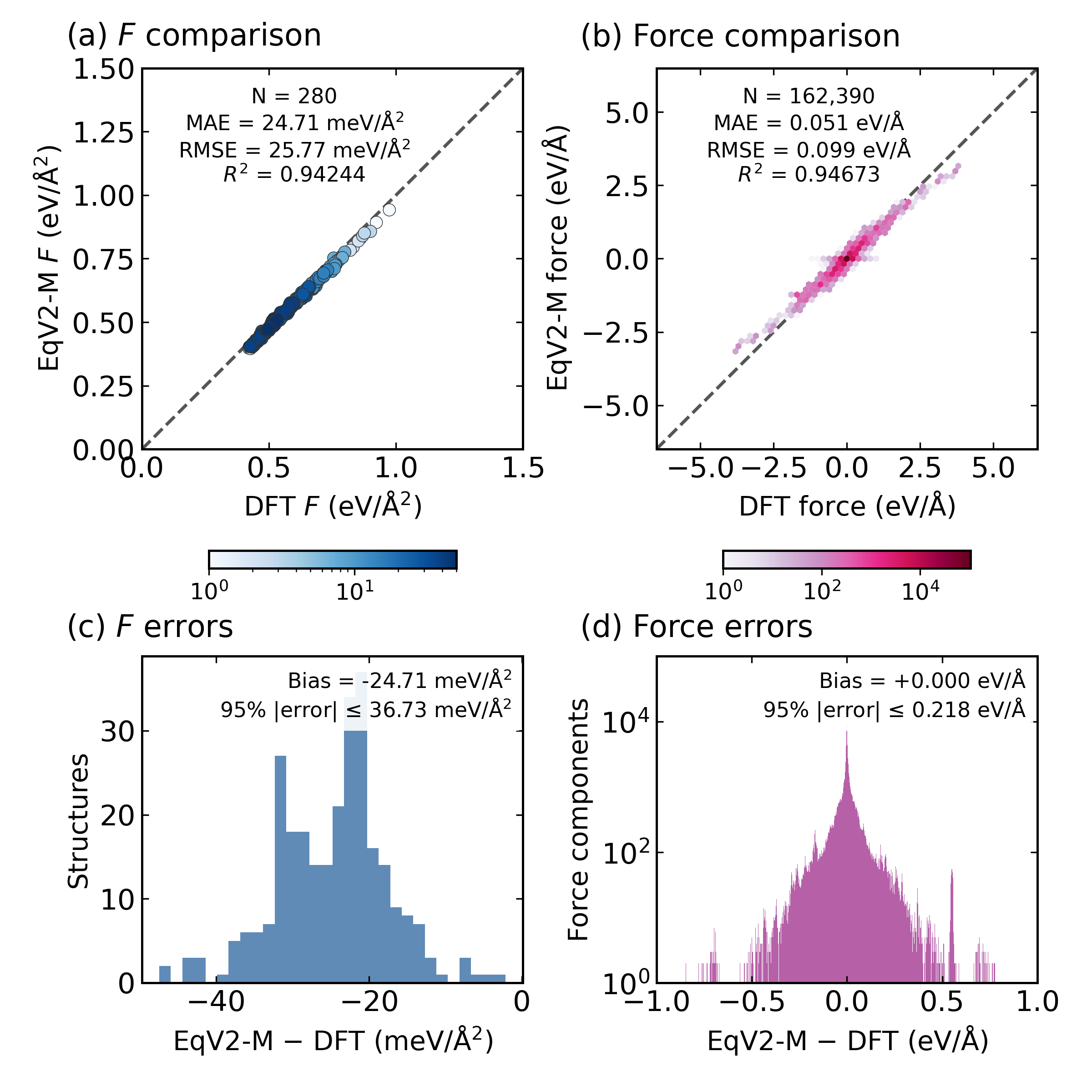}
	\caption{A comparison between universal potential eqV2-M and DFT calculations. (a) Free energy comparison with a MAE of 24.71 meV/\AA$^2$ and RMSE of 25.77 meV/\AA$^2$. The color bar denotes the number of structures. (b) Force comparison with a MAE of 0.051 eV/\AA\ and RMSE of 0.099 eV/\AA. The color bar denotes the number of force components. (c) The distribution of energy errors with a bias at -24.71 meV/\AA$^2$ and 95\% error magnitude within 36.73 meV/\AA$^2$. (d) The distribution of force errors with a bias at 0.000 eV/\AA\ and 95\% error magnitude within 0.218 eV/\AA. The $R^2$ for free energy and forces are 0.999 and 0.947, respectively.}
	\label{fig:validation}
\end{figure}

To identify the most thermodynamically stable structures sampled during the GCMC runs, we first categorized all recorded configurations according to their composition. For each unique combination of $(N_\mathrm{Sr}, N_\mathrm{Ti}, N_\mathrm{O})$, we selected the configuration with the lowest total energy $U$. This procedure yielded 280 unique compositions for the $\Sigma$5(310)[001] SrTiO$_3$ GBs. To assess the accuracy of the universal potential, we conducted DFT calculations by VASP for these 280 structures (details in Supplemental Material) and compared the resulting energies and atomic forces with those predicted by eqV2-M. Since GBs have different compositions and number of atoms, we chose to compare the free energy $F$, instead of the total energy $U$. This introduces a constant shift of each GB structure and so the mean absolute errors (MAE) and Root Mean Square Error (RMSE) would be the same as $U$ but the energy distribution range of $F$ is significantly smaller than $U$. Figure~\ref{fig:validation} compares the DFT free energies per unit area and individual atomic forces with the corresponding eqV2-M predictions. The two methods show good agreement with MAE 24.71 meV/\AA$^2$ for free energy and 0.05 eV/\AA\, for force, respectively. Noticeably, the force error increases when the magnitude of force increases.

\subsubsection{Chemical potential phase diagram}
To evaluate the thermodynamic stability of the obtained 280 GB structures and construct a chemical potential phase diagram, we define the free energy of each structure as a function of the chemical potentials:
		\begin{eqnarray}
			\label{eq:free_energy}
			F=U-(\Delta \mu_{\rm{Sr}}+\mu_{\rm{Sr}}^{0})N_{\rm{Sr}}-(\Delta \mu_{\rm{Ti}}+\mu_{\rm{Ti}}^{0})N_{\rm{Ti}}-(\Delta \mu_{\rm{O}}+\mu_{\rm{O}}^{0})N_{\rm{O}},
		\end{eqnarray}
where $U$ is the total energy of the GB structures evaluated by the uMLIP, $\Delta \mu_i = \mu_i-\mu_i^{0}$ ($i$ = Sr, Ti, or O) is the chemical potential relative to the corresponding elemental reference state, and $N_i$ is the number of atoms of species $i$. For the comparison between uMLIP and DFT in Fig.~\ref{fig:validation}, we chose $\Delta\mu_\mathrm{Sr}$ = -3 eV and $\Delta\mu_\mathrm{Ti}$ = -4 eV, with $\Delta\mu_\mathrm{O}$ determined by the formation energy of SrTiO$_3$.

Because the GB is in equilibrium with the surrounding SrTiO$_{3}$ grains, the chemical potentials must lie within the stability region of SrTiO$_{3}$, while also avoiding precipitation of secondary phases. We therefore impose the following constraints:

\begin{eqnarray}
\label{eq:chemical_potential}
\Delta \mu_{\rm{Sr}}& \le & 0, \\ \nonumber
\Delta \mu_{\rm{Ti}}& \le& 0, \\ \nonumber
\Delta \mu_{\rm{O}}& \le& 0, \\ \nonumber
\Delta \mu_{\rm{Sr}}+\Delta \mu_{\rm{O}}& \le& \Delta H_f(\rm{SrO)}, \\ \nonumber
\Delta \mu_{\rm{Sr}}+2\Delta \mu_{\rm{O}}& \le& \Delta H_f(\rm{SrO_2)}, \\ \nonumber
\Delta \mu_{\rm{Ti}}+\Delta \mu_{\rm{O}}& \le& \Delta H_f(\rm{TiO)}, \\ \nonumber
\Delta \mu_{\rm{Ti}}+2\Delta \mu_{\rm{O}}& \le& \Delta H_f(\rm{TiO_{2})}, \\ \nonumber
2\Delta \mu_{\rm{Ti}}+3\Delta \mu_{\rm{O}}& \le& \Delta H_f(\rm{Ti_{2}O_{3})}, \\ \nonumber
\Delta \mu_{\rm{O}}& =& \{\Delta H_f(\rm{SrTiO_{3}})-\Delta \mu_{\rm{Sr}}-\Delta \mu_{\rm{Ti}}\}/3,
\end{eqnarray}
where $\Delta H_f$ is the formation energy per formula unit, $\Delta H_f$(SrTiO$_{3}$) = -17.87, $\Delta H_f$(SrO) = -6.188, $\Delta H_f$(TiO$_{2}$) = -10.45, and $\Delta H$(Ti$_{2}$O$_{3}$) = -16.60 (eV/f.u.). These values were calculated by eqV2-M and are consistent with the corresponding DFT results (Table~S1). 

\begin{figure}
  \includegraphics[width=\textwidth]{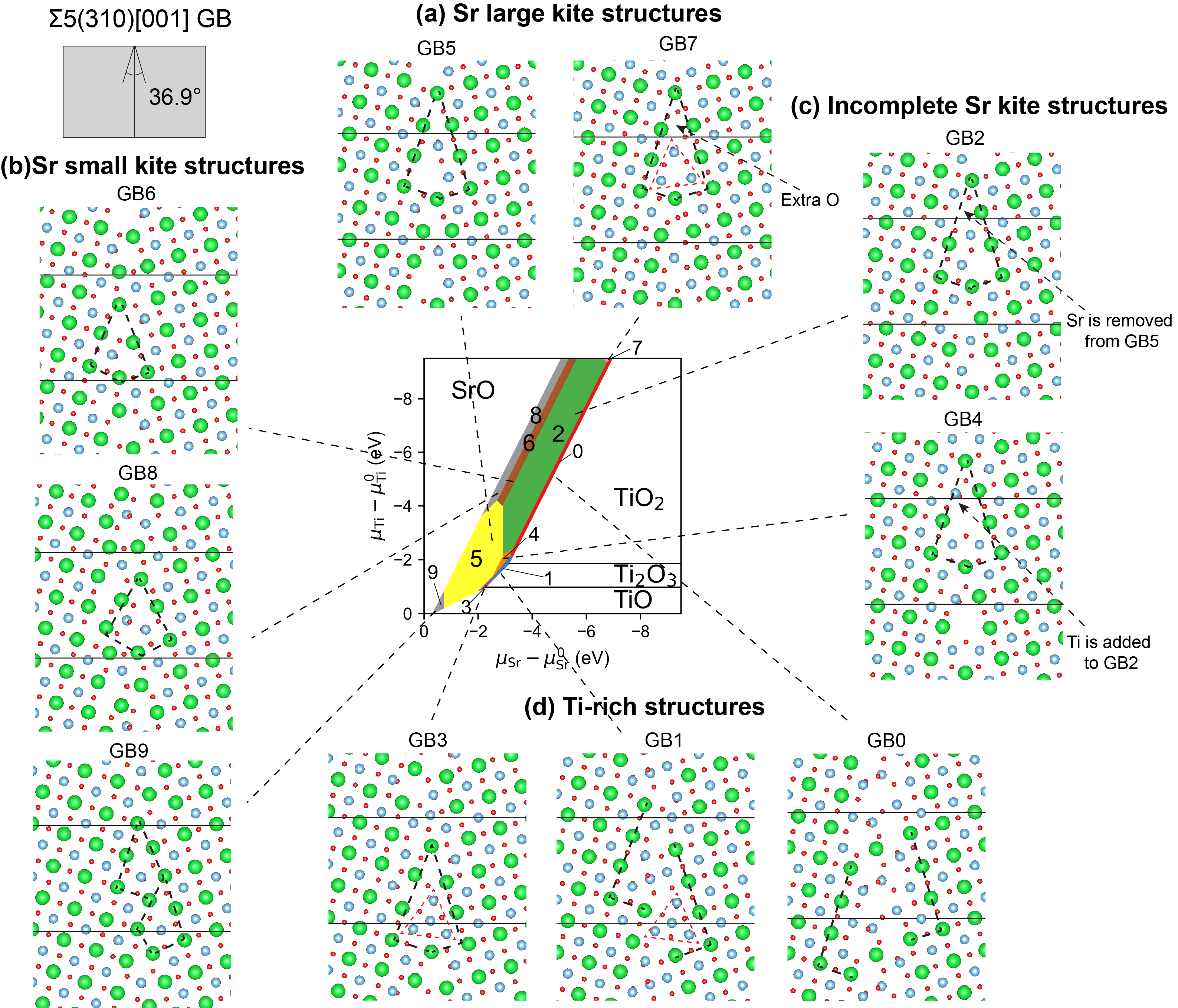}
  \caption{A chemical potential phase diagram of $\Sigma$5(310)[001] SrTiO$_3$ GBs with the corresponding structures. The thin dashed lines are guides to the eye. The thick dashed lines mark the outlines of the GB structures. The thin red dashed lines mark the triangular TiO$_x$ structure. White regions are the forbidden regions where bulk SrO, TiO$_2$, Ti$_2$O$_3$, or TiO precipitate. The chemical potentials below $\Delta \mu_{\rm{Sr}} = \Delta \mu_{\rm{Ti}} = - 9.5$\,eV were cut to avoid GBs with excessive O$_2$ incorporation. The $\Sigma$5(310)[001] GB phase diagram shows a competition between large and small-kite motifs.}
  \label{fig:sigma5_310_001}
\end{figure}

Within the allowed chemical potential region, the thermodynamically stable GB phases are determined by the lower envelope of the free energies of all sampled structures, which is the convex hull in chemical potential space.
The resulting phase diagram for the SrTiO$_3$ $\Sigma$5(310)[001] GB, projected onto the ($\mu_{\rm{Sr}}, \mu_{\rm{Ti}}$) plane, with the corresponding atomic structures of ten stable phases, is shown in Fig.~\ref{fig:sigma5_310_001}, together with forbidden SrO, SrO$_2$, TiO, TiO$_{2}$, and Ti$_{2}$O$_{3}$ precipitation regions. Equivalent projections onto the ($\mu_{\rm{Sr}}, \mu_{\rm{O}}$) or ($\mu_{\rm{Ti}}, \mu_{\rm{O}}$) planes are shown in Fig.~S4. Further details of free energy evaluation, convex hull construction, and phase diagram projections are provided in the Supplemental Material, Section~S2.

As seen, these GB structures are composed of ordered kite-shaped structural units, highlighted by dashed outlines in each subpanel. Based on their common structural motifs, we classify the stable GB phases into four groups:
(a) In the first group (Sr large kite; GB5 and GB7), the GBs feature large kite motifs outlined by eight Sr atoms, with the Sr atoms at the top and bottom shared between adjacent kites within the GB plane.
(b) In the second group (Sr small kite; GB6, GB8, and GB9), the GBs consist of smaller kite motif outlined by six Sr atoms. These phases are stabilized in the chemical potential region close to the SrO precipitation boundaries.
(c) In the third group (incomplete Sr kite; GB2, and GB4), the GBs region is not fully occupied and the large Sr kite motifs are only partially formed. These phases are stabilized near the TiO$_2$ and Ti$_2$O$_3$ precipitation boundaries. 
(d) In the fourth group (Ti-rich; GB0, GB1, and GB3), excess Ti atoms are segregated within the large Sr kite framework. In particular, GB1 and GB3 contain triangular TiO$_x$ motif, outlined by red dashed lines, that are similar to structural motifs observed experimentally \mbox{\cite{wu2022engineering,yang2021determination,browning1999influence}}. The similar triangular motif also appears in GB7, although the center Ti is missing. 
Consistent with this interpretation, the Ti-rich phases are stabilized near the chemical potential region where bulk TiO and Ti$_2$O$_3$ can be precipitated. The explicit compositions of these four groups of GBs evolve across the chemical potential region, and they delicately determine the electronic properties of each GB, as will be discussed below.

\begin{figure}
	\includegraphics[width=0.5\textwidth]{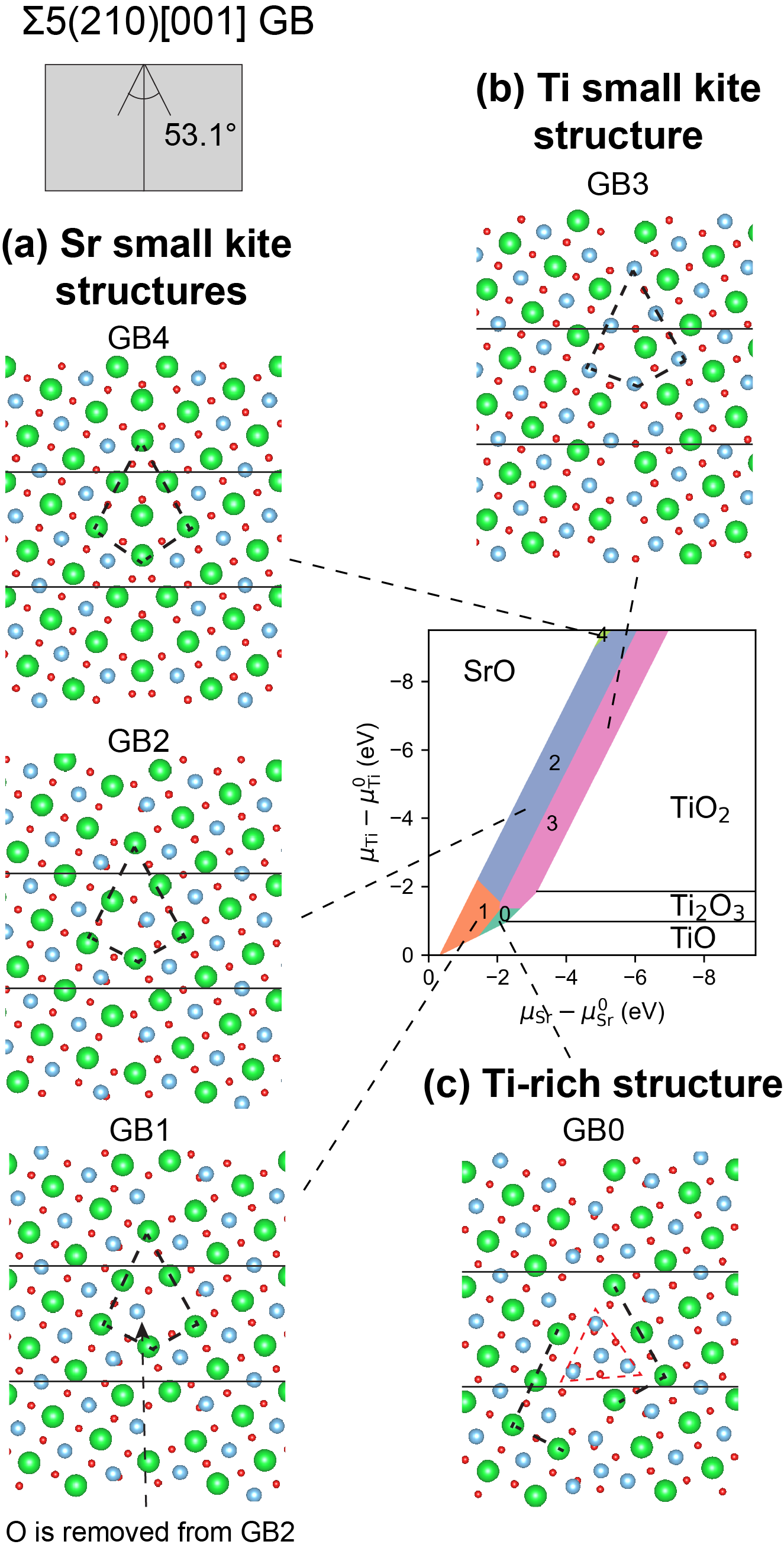}
	\caption{A chemical potential phase diagram of $\Sigma$5(210)[001] SrTiO$_3$ GBs and the corresponding structures. The lines are guides to the eye. The thick dashed lines mark the outlines of the GB structures. The thin red dashed lines mark the triangular TiO$_x$ structure. White regions in the phase diagram are the forbidden regions where bulk SrO, TiO$_2$, Ti$_2$O$_3$, or TiO precipitate. The chemical potentials below $\Delta \mu_{\rm{Sr}} = \Delta \mu_{\rm{Ti}} = - 9.5$\,eV were cut to avoid GBs with excessive O$_2$ incorporation. The $\Sigma$5(210)[001] GB phase diagram is dominated by small-kite motifs.}
	\label{fig:sigma5_210_001}
\end{figure}

\begin{figure}
	\includegraphics[width=0.8\textwidth]{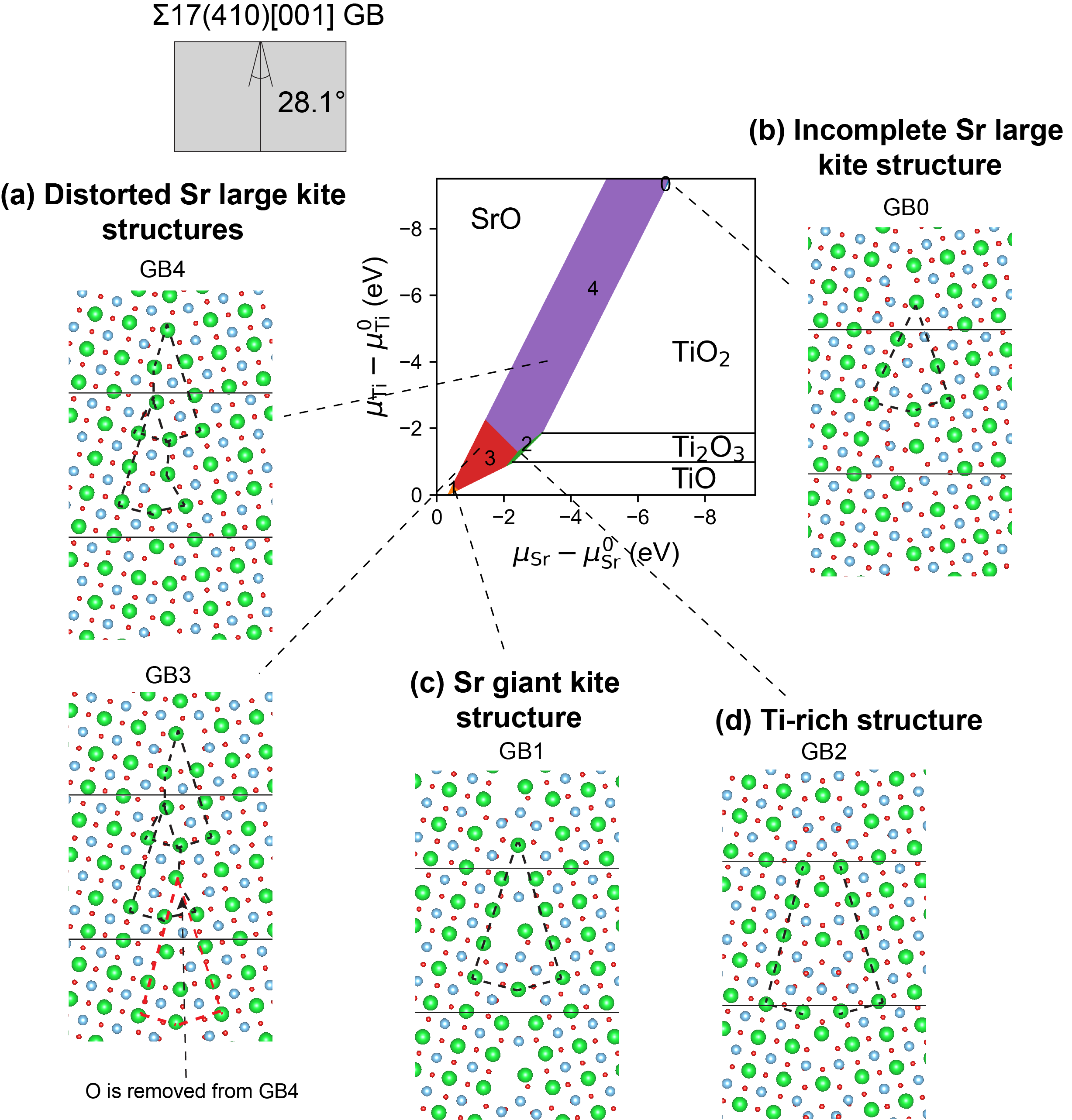}
	\caption{A chemical potential phase diagram of $\Sigma$17(410)[001] SrTiO$_3$ GBs (GB) and the corresponding structures. The thin dashed line are guides to the eye. The thick dashed lines mark the outlines of GB structures. The red dashed lines mark distorted giant Sr-kite structure. White regions are the forbidden regions where bulk SrO, TiO$_2$, Ti$_2$O$_3$, or TiO precipitate. The chemical potentials below $\Delta \mu_{\rm{Sr}} = \Delta \mu_{\rm{Ti}} = - 9.5$\,eV were cut to avoid GBs with excessive O$_2$ incorporation. The $\Sigma$17(410)[001] GB phase diagram is dominated by large-kite motifs.}
	\label{fig:sigma17_410_001}
\end{figure}

Similarly, we studied $\Sigma$5(210)[001] and $\Sigma$17(410)[001] GBs, to investigate the effects of the GB plane and misorientation angle. $\Sigma$5(210)[001] GB shares the same misorientation angle with $\Sigma$5(310)[001] GB but has a different GB plane and a large inclination angle, whereas $\Sigma$17(410)[001] GB has different misorientation angle, different GB plane, and a smaller inclination angle, as illustrated in Fig.~S10 and ~S11.
The constructed phase diagrams and corresponding structures of $\Sigma$5(210)[001] and $\Sigma$17(410)[001] SrTiO$_3$ GBs are shown in Figures~\ref{fig:sigma5_210_001} and ~\ref{fig:sigma17_410_001}, respectively. Both sets of GBs show simpler phase behaviors compared to $\Sigma$5(310)[001] GBs. 
For the $\Sigma$5(210)[001] (Fig.~\ref{fig:sigma5_210_001}), small kite structures occupy a large fraction of the phase diagram. GB1, GB2, and GB4 adopt Sr-based small kite motifs, outlined by six Sr atoms, and are stabilized near the SrO precipitation boundary. GB3 forms a Ti-based small kite structure, outlined by six Ti atoms, and is near the TiO$_2$ precipitation region. GB0 is more strongly Ti-rich, with excess Ti segregated between Sr columns and a distinct TiO$_x$ motif present at the GB core. This TiO$_x$ motif is similar to those observed in $\Sigma5$(310)[001] GB1 and GB3. Sr small kite structures with partial Ti and O segregation have been observed in experiments \cite{lee2011effect}, and closely resemble our predicted GB1 and GB2 structures.

For the $\Sigma$17(410)[001] GBs (Fig.~\ref{fig:sigma17_410_001}), distorted Sr large kites, outlined by eight Sr atoms, occupy a substantial portion of the phase diagram. GB3 and GB4 exhibit very similar distorted Sr large-kite structures, with GB4 containing one extra oxygen. GB0 forms an incomplete Sr large kite structure where one Sr in the large kite outline is replaced by Ti. At the tip of the chemical potential phase diagram, GB1 adopts a complete Sr giant kite structure outlined by nine Sr atoms, with the Sr atoms at the top and bottom shared between adjacent giant kites within the GB plane. Near the TiO and Ti$_2$O$_3$ precipitation regions, GB2 forms a Ti-rich structure, with excessive Ti atoms enclosed within the giant Sr kite outline. In experiments, giant Sr kite structures with partially occupied Sr and O have been observed \cite{browning1995observation}, closely resembling the GB1, GB3, and GB4 predicted here. In particular, GB3 and GB4 can also be viewed as strongly distorted giant Sr-kite structures in which several Sr atoms deviate from the ideal kite outline, as indicated by the red dashed lines in Fig.~\ref{fig:sigma17_410_001}(a).

The observations that  $\Sigma$5(210)[001] and $\Sigma$17(410)[001] are dominated by either small kite or large kite structures, respectively, whereas $\Sigma$5(310)[001] accommodates both motif sizes, suggest that GBs with an intermediate GB plane inclination can be stabilized by competing structural units characteristic of the two limiting inclinations. This competition of small and large kites partitions the $\Sigma5(310)[001]$ chemical potential phase diagram into multiple stability regions associated with distinct GB motifs.

To further assess the robustness of our GCMC predicted GB structures, we performed several additional calculations. First, we performed GCMC simulations starting from GB configurations with a larger initial gap of 10\,\AA\  between the two slabs. This setup tests whether the method can reconstruct bulk-like regions adjacent to the interface and subsequently fill the interfacial gap with stable GB structures. Second, we performed GCMC calculations using wider GB supercells with twice the interface area. Both tests were carried out for $\Sigma5(210)[001]$ and $\Sigma5(310)[001]$ GBs, and the resulting phase diagrams are shown in the Supplemental Material Figures.~S6, S7, S8, and S9. For both the large-gap and wider-interface calculations, the predicted phase diagrams and GB structures are comparable with those in Fig.~\ref{fig:sigma5_310_001} and Fig.~\ref{fig:sigma5_210_001}, further supporting the robustness of the GCMC methodology. Additional details of these tests are provided in Supplemental Material, Section~S4.
Third, we applied the same GCMC method to asymmetric $\Sigma5(430)//(100)[001]$ GBs. The resulting phase diagram contains both small kites and large kite structures, as summarized in Supplemental Materials, Section~S5. We note that the $\Sigma5(430)//(100)[001]$ GB shares the same misorientation angle and CSL index with $\Sigma5(210)[001]$ and $\Sigma5(310)[001]$ GBs, but differs in GB plane and exhibits an asymmetric inclination geometry, as illustrated in Fig.~S10. In experiments, for $\Sigma5[001]$ GBs with a misorientation angle of $36.8^\circ$, the GB plane can vary among $(210)$, $(310)$, and $(430)//(100)$, resulting in a mixture of different boundary-plane configurations \cite{lee2011effect,rybak2025atomic}. 
Finally, we repeated GCMC calculations for the $\Sigma5(210)[001]$ GB using two additional uMLIPs: the MatterSim-5M model developed by Microsoft \cite{yang2024mattersim} and the NequIP-OAM-M model \cite{tan2026high,kavanagh2026fast} developed by the Materials Intelligence Group at Harvard University. The resulting phase diagrams, shown in the Supplemental Material Figures.~S14 and ~S15, exhibit slight differences due to the different formation energies among the models (Tables.~S3 and ~S4), whereas the corresponding GB structural motifs remain comparable with those predicted by eqV2-M.

These results demonstrate that GCMC sampling with a uMLIP can efficiently identify stable SrTiO$_3$ GB structures across a range of chemical potentials and boundary geometries. Beyond reproducing experimentally observed cation motifs, the approach resolves oxygen positions and stoichiometry, providing a structural basis for investigating the mechanical, dielectric, and electronic properties of perovskite GBs.


\subsubsection{Electronic structure}
Following the construction of the phase diagrams for the stable GB structures, we further performed DFT calculations to investigate the electronic and electrostatic properties of the ten $\Sigma$5(310)[001] GBs. For each GB, we first construct a periodic boundary condition (p.b.c) supercell of each GB containing two equivalent GBs. Then the structures were initially relaxed by eqV2-M, allowing both the atomic positions and supercell dimensions to vary so that local polarization within the grains could develop. The structures were then further relaxed using DFT, with the atomic positions additionally optimized. Additional details of constructing the p.b.c supercell are provided in the Supplemental Material, Section~S7.

\begin{figure}
	\includegraphics[width=0.95\textwidth]{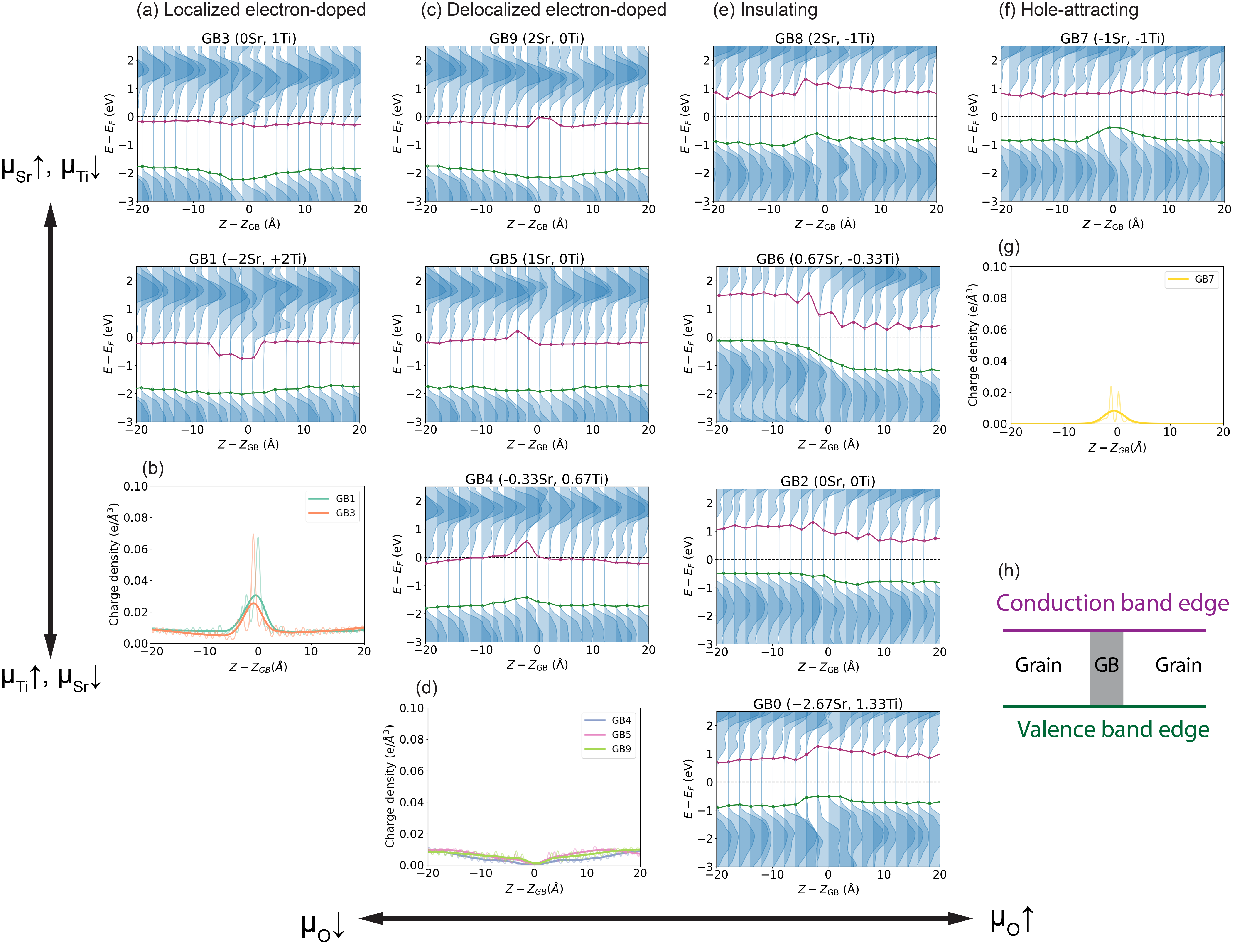}
	\caption{Electronic structures of $\Sigma$5(310)[001] SrTiO$_3$ GBs. (a) PDOS of localized electron-doped GBs: GB3 and GB1. (b) Partial electron density integrated over the energy range from -0.8 eV to 0.4 eV relative to $E_F$. (c) PDOS of delocalized electron-doped GBs: GB9, GB5, and GB4. (d) Partial electron density integrated over the energy range from -0.8 eV to 0.4 eV relative to $E_F$. (e) PDOS of insulating GBs that form $p-n$ junction-like band structures, including GB8, GB6, GB2, and GB0. (f) PDOS of hole-attracting GB7. (g) Electron density integrated over the energy range from -1.0 eV to 0 eV relative to $E_F$. (h) Schematic illustrates the GB and grain regions together with the conduction- and valence-band edges, using colors and spatial arrangements corresponding to the PDOS panels. In each PDOS panel, the Fermi level $E_F$ is indicated by the black dashed lines. In partial charge density profiles, the thin curves represent the $x-y$ planar-averaged charge density while the thick curves represent the macroscopically averaged electron density.  Across the figure, $\mu_{\rm{O}}$ increases from left to right, while $\mu_{\rm{Sr}}$ ($\mu_{\rm{Ti}}$) increases from bottom (top) to the top (bottom). $\Sigma$5(310)[001] SrTiO$_3$ GBs form four distinct electronic regimes.}
	\label{fig:GB_pDOS}
\end{figure}

To examine how grain-boundary composition and structure affect the electronic structure, we calculated the atomic-layer projected density of states (PDOS) along the direction normal to the GB plane. For each structure, the supercell was divided into 2~\AA\ wide layers along the $\hat z$ direction, normal to the GB plane, and DOS was projected onto the atoms of each layer. The PDOS for GB0 -- GB9 of the $\Sigma$5(310)[001] GBs are shown in Figure~\ref{fig:GB_pDOS}. All energies are referenced to the Fermi level $E_F$ of each GB, indicated by the black dashed line in each panel. We use purple lines and dark green lines to mark the conduction and valence band edge, as schematically plotted in Fig.~\ref{fig:GB_pDOS} (h).
Based on PDOS, $\Sigma$5(310)[001] GBs can be classified into four regimes, arranged in columns with increasing $\mu_{\rm O}$: (a) localized electron-doped, (b) delocalized electron-doped, (c) insulating with a  $p-n$ junction-like profile, and (d) hole-attracting.
To quantify the composition, we define the excess Sr and Ti per GB interface area relative to stoichiometric SrTiO$_3$ using O as the reference, $(N_{\rm{Sr}}-N_{\rm{O}}/3, N_{\rm{Ti}} - N_{\rm{O}}/3)$, and labeled them in each panel of Fig.~\ref{fig:GB_pDOS}. The periodic area of the $\Sigma5$(310)[001] GB  is $\sqrt{10}\times3.9126^2$\AA$^2$ = 48.41\,\AA$^2$ with small variations among the relaxed GB structures.

In the localized electron-doped regime (Fig.~\ref{fig:GB_pDOS}(a)), GB3 (0 Sr, 1 Ti) and GB1 (-2 Sr, 2 Ti) contain extra cationic charge and therefore carry positive nominal GB-core charge when evaluated using the formal valences Sr$^{2+}$, Ti$^{4+}$ and O$^{2-}$. These excess cations donate electrons to Ti-$3d$ conduction band states, raising $E_F$ into the conduction band and producing electron accumulation near the GB. The excess electron density is predominantly concentrated around the GB core though with minor distribution in the grain region, as evident by both the layer-resolved PDOS and the partial charge density integrated from -0.8 to +0.4 eV relative to $E_F$ (Fig.~\ref{fig:GB_pDOS}(b)). The charge density is in units of $e$/\AA$^3$. Integrating the charge density over the length of the supercell and multiplying it by the GB area leads to the total charge per supercell in the energy integration window.
In contrast, in the delocalized electron-doped regime (Fig.~\ref{fig:GB_pDOS}(c)), GB9 (2 Sr, 0 Ti), GB5 (1Sr, 0Ti), and GB4 (-0.33 Sr, 0.67 Ti) also contain excess cations, but the donated electrons are not strongly confined to the GB core and instead extend into the neighboring grain regions, as seen in partial charge density (integrated from -0.8 to +0.4 eV relative to $E_F$) in Fig.~\ref{fig:GB_pDOS}(d). 
In the insulating regime (Fig.~\ref{fig:GB_pDOS}(e)), GB8 (2Sr, -1Ti), GB6 (0.67 Sr, -0.33 Ti), GB2 (0Sr, 0Ti), and GB0 (-2.67 Sr, 1.33 Ti) are nominally charge neutral based on formal valences Sr$^{2+}$ and Ti$^{4+}$, and their $E_F$ remain within the band gap. Despite the absence of occupied conduction states, several of these GBs, particularly GB2 and GB6, are structurally asymmetric. The associated interfacial polarization and electrostatic potential offset shift the local band edges on the two sides of the GB, producing a $p-n$ junction-like band alignment profile in the layer-resolved PDOS even though no free electron or hole carriers are present.
Finally, in the hole-attracting regime (Fig.~\ref{fig:GB_pDOS}(f)), GB7 (-1 Sr, -1 Ti) is O-rich and develops additional O-$2p$ derived states near the valence-band edge. Combined with the upward shift of the local valence-band edge, these states produce a band alignment that energetically favors hole accumulation at the GB. The partial charge density integrating from -1 to 0 eV relative to $E_F$ shows that the corresponding valence-edge states are predominantly localized at the GB core (Fig.~\ref{fig:GB_pDOS}(g)).

\begin{figure}
	\includegraphics[width=\textwidth]{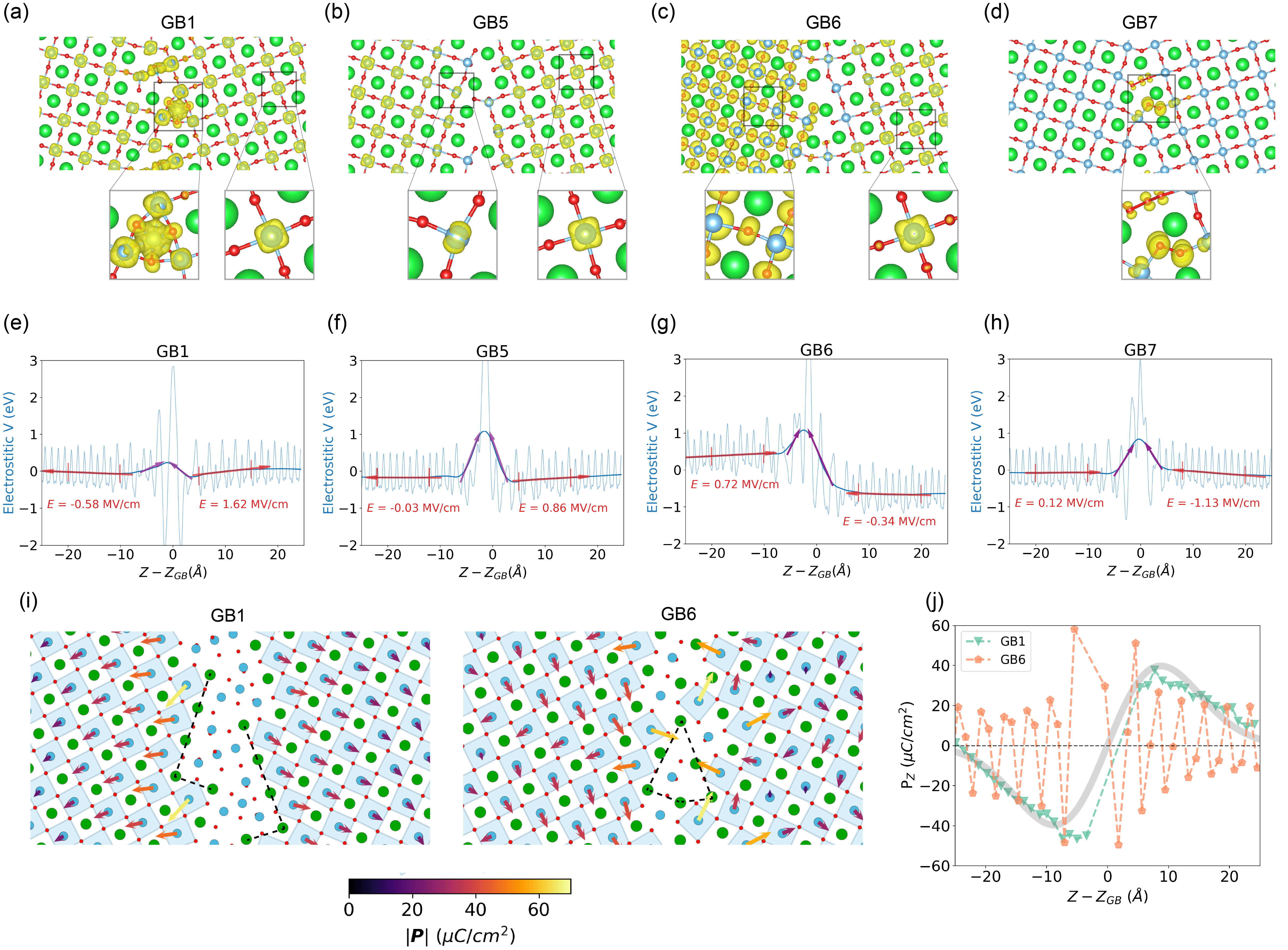}
	\caption{(a--d) Partial charge density distributions for GB1, GB5, GB6, and GB7, integrated over energy windows of [-0.8, 0.4], [-0.8, 0.4], [-1, 1], and [-1, 0] eV relative to $E_F$. The isosurfaces correspond to an electron density of 0.004 e/\AA$^3$. The insets show the enlarged view of indicated regions of charge density, as marked by the squares. (e--h) Corresponding electrostatic potential profiles. The thin blue curves represent the $x-y$ planar-averaged electron electrostatic potential while the thick blue curves represent the macroscopically averaged electron electrostatic potential. Short vertical red lines mark the region outside the GB core for the linear fit, and the red arrows are the fits, whose slopes correspond to the local electric field. The purple arrows schematize the electric field direction inside the GB core. The negative field points toward left and the positive field points toward right. (i) Local unit-cell polarization vectors near GB1 and GB6. (j) Polarization component $P_z$ normal to the GB plane for GB1 and GB6. The thick gray line represents a fit to the GB1 profile using the first derivative of a Gaussian.}
	\label{fig:GB_polarization}
\end{figure}

To understand the origin of the electronic structure of each regime, we analyzed the partial charge density near $E_F$. One representative GB of each regime is selected, and the corresponding partial charge densities are shown in Figure~\ref{fig:GB_polarization} (a--d). For GB1 (Fig.~\ref{fig:GB_polarization}(a)), charge density was integrated from -0.8 to 0.4 eV relative to $E_F$, and the resulting charge is concentrated primarily at the GB core, especially around the excess Ti atoms. At the GB core, the charge density is elongated along the Ti--O directions, whereas in the grain region the rest of the charge density is centered on Ti sites but is elongated toward the Ti--Sr directions, as illustrated in the zoomed-in plots. This contrast suggests that the lower-coordination TiO$_x$ environment at the GB core promotes stronger mixing between the Ti-3$d$ $t_{2g}$ and $e_g$ orbitals, while in the grain region the conventional STO crystal-field splitting is largely preserved, such that the $t_{2g}$ states remain the lowest-energy Ti-$3d$ states and are occupied first. This $t_{2g}$ and $e_g$ mixture in the STO GB region has been observed in electron energy loss spectroscopy (EELS) experiments, as evidenced by the broadening and partial merging of the two characteristic peaks \cite{yang2021determination,liao2021nanoscale}.
Indeed, the TiO$_x$ structures observed in GB1 and GB3 are crucial to the Ti-$3d$ hybridization at the GB core; this hybridization is not a simple consequence of stoichiometry and nominal charge alone. As evidence, the $\Sigma5(210)[001]$ GB0, which contains the same TiO$_x$ motif, shows a similar localized electron-doped behavior as well, as shown in Fig.~S19.
By contrast, in GB5 (Fig.~\ref{fig:GB_polarization} (b)), although excess cations are also present at the GB core, the local structure does not produce similarly strong Ti-$3d$ orbital mixing. As a result, the donated electrons are less confined to the GB core and instead preferentially occupy the more extended Ti-$3d$ $t_{2g}$-like states in the neighboring grain regions, consistent with the more delocalized electron-doping behavior (See insets in Fig.~\ref{fig:GB_polarization}(b)).
In the insulating cases (Fig.~\ref{fig:GB_polarization}(c)), the charge density integrated from -1 to 1 eV relative to $E_F$ highlights the valence band edge on the one side of the GB and conduction band edge on the other, reflecting the strong $p-n$ junction like band bending. The valence band edge is dominated by O-$2p$ states, showing $p_{\pi}$ features extending perpendicular to the Ti-O bonds, whereas the conduction band edge is composed mainly of Ti-$3d$ $t_{2g}$ like states. For the hole-attracting case, the partial charge density integrated from -1 to 0 eV relative to $E_F$ (Fig.~\ref{fig:GB_polarization}(d)), shows that the valence edge states are strongly localized at the GB core and originate primarily from O-$2p$-derived states with hybridization with Ti orbitals, originated from the oxygen-rich configuration.

\subsubsection{Electrostatics and local polarizations}

Next, we examine how the GB structure and the associated charge redistribution collectively determine the local electrostatics. The DFT-calculated electron electrostatic potentials and their macroscopic averages for GB1, GB5, GB6, and GB7 are shown in Fig.~\ref{fig:GB_polarization} (e--h), which are defined as $V = -e\phi$,  where $\phi$ is the conventional electrostatic potential, so that the electric field is given by $E_z=(1/e)\,dV/dz$.
In the bulk-like grain region, atomic layers parallel to $(310)$ plane alternate between Sr--Ti--O and O--O layers with nominal charges of +4 and -4, respectively. The microscopic potential (thin blue curves) therefore exhibits strong oscillations, whereas the macroscopically averaged potential (thick blue curves), obtained by convolution with a Gaussian window function, becomes smooth in the grain interior (thick blue lines). 
At the GB core (about -5\,\AA\ $< Z-Z_{GB} < 5$\,\AA), however, changes in local atomic density and bonding produce a pronounced hump in all GB structures, irrespective of their electronic band structures.
The corresponding macroscopic potential generally exhibits a positive slope on the left and a negative slope on the right sides, producing a head-to-head electric-field configuration within the GB core, as illustrated by the purple arrows. This short-range variation is associated with the mean inner potential (MIP) contribution of the reconstructed GB core and extends only over a few unit cells, on the order of $\approx$1 nm \mbox{\cite{toyama2024direct}}.

Superimposed on this short-range GB core potential is a more slowly varying electrostatic potential associated with space charge. The regions immediately outside the GB core, indicated by the short vertical red bars in Fig.~\ref{fig:GB_polarization}(e--h) , are fitted with linear functions (red arrows), whose slopes give the electric fields on either side of the GB. For GB1 (Fig.~\ref{fig:GB_polarization}(e)), the Ti-rich GB core and redistribution of donated conduction-band electrons produce a negative field on the left and a positive field on the right,  corresponding to a tail-to-tail configuration. Thus, the short-range electric field within the GB core and the longer-range field outside the core point toward opposite directions. The same qualitative behavior persists in GB5 (Fig.~\ref{fig:GB_polarization}(f)), although outer electric field is much weaker. 

By contrast, GB6 (Fig.~\ref{fig:GB_polarization}(g)) has an almost flat potential outside the core and only a weak head-to-head field, aligned with the short-range core field. GB7 (Fig.~\ref{fig:GB_polarization}(h)), in which excess oxygen gives rise to a negatively charged GB region, likewise exhibits a head-to-head electric-field pattern outside the GB core. We note that the electrostatic potential and electric-field profiles are generally asymmetric because GB structures lack mirror symmetry about the GB plane. This asymmetry is particularly pronounced in GB6, where the unequal electron electrostatic potential on the two sides of the GB produces a $p-n$ junction like profile consistent with the band alignment in Fig.~\ref{fig:GB_pDOS} (e).

Tail-to-tail electric field outside Ti-rich STO GBs have been observed experimentally \cite{yang2021determination,gilgenbach2025measurement,wang2002determination}, with the associated space-charge region extending from several to tens of nanometers depending on the carrier concentration and screening conditions. Conversely, head-to-head electric fields associated with negatively charged STO GBs have also been observed experimentally \cite{ravikumar1995direct}. These results highlight two characteristic electrostatic length scales at STO GBs: a short-range GB core contribution governed by the reconstructed atomic structure, and a longer range space-charge contribution governed by the GB charge and its electrostatic screening.

From the electric fields, we can further obtain the total charge density through Poisson's equation, $\rho_\mathrm{total} = \epsilon_0\frac{dE}{dz}$, where $\epsilon_0 = 8.854\times10^{-12}$ F/m is the vacuum permittivity (Supplemental Material Sec.~S11). Within the GB core, the head-to-head electric-field pattern gives rise to a characteristic positive--negative--positive charge-density profile. The positive and negative contributions do not necessarily cancel upon integration over the entire GB core, so the net integrated charge can be positive, negative, or approximately zero. Its sign is determined by $\int_{z_1}^{z_2}\rho_\mathrm{total}dz = \epsilon_0[E(z_2)-E(z_1)]$, where $z_1$ and $z_2$ denote the left and right boundaries of the GB-core region, respectively. 
For GB1, the field is negative on the left and positive on the right ($E(z_2)-E(z_1)>0$), yielding a net positive GB core-charge.
In contrast, GB7 exhibits a positive field on the left and a negative field on the right ($E(z_2)-E(z_1)<0$), yielding a net negative GB-core charge. For GB5 and GB6, the field difference across the core is small, and the integrated charge is close to zero. The electric-field and total charge density profiles of GB1 and GB7 are summarized in Fig.~S22. 
Thus, the local electrostatics are collectively determined by the atomic-scale charge distribution, the induced polarization, and the redistribution of donated or depleted electronic carriers, all of which depend sensitively on GB structure and composition.

The polarization response provides an additional contribution to this local charge redistribution. In cation-rich GBs, local structural distortions can generate tail-to-tail polarization patterns and associated negative bound charge near the GB core, partially screening its positive charge. We estimate the local polarization of each identifiable perovskite unit cell by summing the atomic displacements relative to a locally constructed centrosymmetric cubic unit-cell reference, multiplied by the corresponding Born effective charge tensors and normalized by the cell volume. The resulting polarization patterns of GB1 and GB6 are compared in Fig.~\ref{fig:GB_polarization}(i). In Ti-rich GB1, the polarization vectors predominantly point away from the boundary, forming a clear tail-to-tail pattern, whereas GB6 shows weaker, less correlated polar distortions without a well-defined collective direction.
Fig.~\ref{fig:GB_polarization}(j) shows the polarization component $P_z$ normal to the GB plane. In GB1, $P_z$ changes sign across the boundary, reaches its largest magnitude immediately outside the core, and decays toward zero into the grains. GB6 instead exhibits sign fluctuations without a systematic spatial pattern. The GB1 profile is well described by the first derivative of a Gaussian, as indicated by the thick gray line. Through $\rho_{\mathrm{bound}}=-dP_z/dz$, this profile produces negative bound charge $\rho_{\mathrm{bound}}$ at the GB core and compensating positive bound charge farther into the grains. 
These results show that the Ti-rich GB1 supports a pronounced tail-to-tail polar response, whereas the charge-neutral GB6 exhibits no clear collective polarization pattern. The local polar response is therefore strongly governed by GB composition and atomic structure, through both reconstructed structural distortion and the local electric fields generated.

We note that these finite p.b.c calculations suggest that cation-rich GBs can favor electron accumulation near the GB core while also supporting more delocalized carriers in the grain region. In experiments with much larger grains, the lower average donor concentration may instead allow weakly conducting or insulating grain interiors to coexist with electron accumulation or trapping near the GB, depending on the self-consistent band bending and occupation of GB-derived states. At the continuum level, this behavior is commonly described by a double-Schottky-barrier model, in which a charged GB core is compensated by space-charge regions extending into the adjoining grains, extending over a few to tens of nanometers \cite{de2003electrical}. As an illustrative case, a Poisson-Boltzmann model treating one excess Ti$^{4+}$ per GB repeat unit as an effective positive sheet charge is presented in Supplemental Materials, Section~S12, while a quantitative description would require self-consistent DFT-continuum treatment \cite{sinai2015multiscale}.

Nevertheless, our DFT results highlight the functional diversity enabled by GB composition and atomic structure. Ti-rich GB1 exhibits donor-like behavior, electron accumulation, and local polar distortions, whereas GBs stabilized at higher $\mu_{\mathrm{O}}$ tend to be insulating or hole-attracting. 
Ti-rich GBs may therefore offer opportunities to couple electron accumulation with local polarization and interfacial magnetism from occupied-Ti \cite{najev2025electronic,shimada2015multiferroic}. More oxidizing conditions may instead favor GB structures that reduce electronic leakage. Structurally asymmetric GBs offer another possibility generating interfacial potential offsets and a band alignment that favors electrons and holes on opposite sides of the boundary, analogous to polarization-driven carrier separation at polar boundaries in halide perovskites \cite{grieder2025carrier}. These results identify chemical potential control, particularly through oxygen pressure, as a practical route to engineering GB electronic states, local polarization, and interfacial electrostatics.

\subsection{Conclusion}
In summary, we adapted a uMLIP-accelerated GCMC framework to investigate the thermodynamic stability of $\Sigma$5(310)[001] SrTiO$_3$ GBs and construct their chemical potential phase diagram. We identify four motif families: large Sr kite structures, small Sr kite structures, incomplete Sr large kite structures, and Ti-rich structures, bounded by the precipitation regions of SrO, TiO$_2$, Ti$_2$O$_3$ and TiO phases. Extending the GCMC sampling to $\Sigma$5(210)[001] and $\Sigma$17(410)[001] further demonstrates that GB geometry strongly influences the competition among those motifs: the larger inclination angle $\Sigma$5(210)[001] GB is dominated by small-kite motifs whereas the lower inclination angle $\Sigma$17(410)[001] GB is dominated by large-kite motifs. Using DFT calculations, we further uncover the local electronic properties of the GCMC-predicted GB phases, revealing four distinct electronic regimes: localized electron-doped, delocalized electron-doped, insulating, and hole-attracting states. The corresponding electrostatic analysis further reveals that the short-range head-to-head GB core field is superimposed on longer-range potential modulations associated with GB charge and carrier redistribution, producing either tail-to-tail or head-to-head field configurations immediately outside the GB core. In Ti-rich GBs, the associated electrostatic response can also induce pronounced tail-to-tail local polarization, generating bound charge that partially screens the GB-core charge. Overall, this study establishes uMLIP-accelerated GCMC as a general framework for efficiently exploring the structure and compositional phase space of oxide perovskite GBs while uncovering how GB stoichiometry and structural motifs couple to electronic reconstruction, electrostatics, and local polarization. By linking thermodynamically accessible GB structures across chemical potential space to their resulting electronic and electrostatic properties, this approach provides a route toward predictive GB engineering in complex oxides and other functional materials.

\section{Acknowledgments}
We acknowledge the helpful discussion with Dr. Tan Zhang.
K. M. was supported by JSPS KAKENHI grant numbers 21J10412 and Overseas Research Fellowships hosted by JSPS and Office of Naval Research. X. H. and A. M. R. were supported by Office of Naval Research, under grant number
N00014-24-1-250. Computational support was provided by the National Energy Research Scientific Computing Center
(NERSC), a U.S. Department of Energy, Office of Science User Facility located at Lawrence Berkeley
National Laboratory, operated under Contract No. DE-AC02-05CH11231.

\bibliography{references}
\end{document}